\documentclass[12pt,thmsa]{article}

\begin{document}

\author{A.I.VOLOKITIN$^{1,2}$ and B.N.J.PERSSON$^1$ \\
1. Institute f\"ur Festk\"orperforschung,Forschungszentrum\\
J\"ulich, D-52425 J\"ulich, Germany \\
2. Samara State Technical University, 443010 Samara,\\
Russia}
\title{Theory of friction: contribution from fluctuating electromagnetic field}
\maketitle

\begin{abstract}
We calculate the friction force between two semi-infinite solids in relative
parallel motion (velocity $V$), and separated by a vacuum gap of width $d$.
The friction force result from coupling via a fluctuating electromagnetic
field, and can be considered as the dissipative part of the van der Waals
interaction. We consider the dependence of the friction force on the
temperature $T$, and present a detailed discussion of the limiting cases of
small and large $V$ and $d$.
\end{abstract}

\section{Introduction}

Because of its great practical importance and because of the development of
new experimental techniques, sliding friction has become a topic of
increasing attention \cite{one}. In this paper we consider the friction
force between two solids in relative motion, separated by a vacuum gap of
width $d$. This ``vacuum'' friction is in most cases of no direct practical
importance since the main contribution to the friction force when a body is
slid on another body comes from the area of real (atomic) contact\cite{one}.
Thus, the frictional stress between two semi-infinite metallic (e.g.,
copper) bodies, moving parallel to each other with the relative velocity $%
V=1\,\mathrm{m/s}$, and separated by the distance $d=10\,\mathrm{\AA }$, is
only (see Ref. \cite{two} and below) $\sim 10^{-6}\,\mathrm{N/m}^{\mathrm{2}%
} $. This stress is extremely small compared with the typical frictional
stress $\sim 10^8\,\mathrm{N/m}^2$ occurring in the area of atomic contact
even for (boundary) lubricated surfaces. Nevertheless, vacuum friction is
important in some special cases (see Ref. \cite{two}), and determines the
ultimate limit to which friction can be reduced. Quantum and thermal
fluctuations of the polarization and the magnetization of solids give rise a
fluctuating electromagnetic field. For two stationary solids the interaction
mediated by this field result in the well-known attractive van der Waals
force. For two solids in relative motion this interaction will also give
rise to a friction force between the bodies. The static aspect of the van
der Waals interaction is well understood but there are still controversial
results concerning the dynamical part. Different author's have recently
studied the van der Waals friction using different approaches, and obtained
results which are in sharp contradiction to each other. The first
calculation of van der Waals friction was done by Teodorovitch\cite{three}.
Schaich and Harris\cite{four}, and Pendry\cite{five} argue that Teodorovitch
calculation is in error. For two metallic bodies Schaich and Harris found
that the friction force is independent of any metal property, in contrast to
the results of other authors. The friction forces calculated by Levitov\cite
{six}, Polevoi\cite{seven} and Mkrtchiau\cite{eight} vanish in the
non-retarded limit (formally obtained when the light velocity $c\rightarrow
\infty $). This result is very surprising (and in our opinion incorrect),
since neglecting retardation is a good approximation at short separations $d$
between the surfaces, in which case one would expect the friction force to
be particularly large. Even at large separation, where the non-retarded
contribution is negligible, our result differs form those of Levitov,
Polevoi and Mkrtchiau. Pendry\cite{five} considered only the case of zero
temperature, and Persson and Zhang\cite{two} the case of small sliding
velocities, and both groups neglected retardation effects. To clarify the
situation we present a straightforward calculation of the van der Waals
friction based on the general theory of the fluctuating field developed by
Rytov\cite{nine} and applied by Lifshitz\cite{ten} for studying the
conservative part of the van der Waals interaction. In this approach the
interaction between the two bodies is mediated by the fluctuating
electromagnetic field which is always present in the vicinity of any
collection of atoms. Beyond the boundaries of a solid this field consist
partly of traveling waves and partly of evanescent waves which are damped
exponentially with the distance away from the surface of the body. The
method we use for calculating the interaction forces is quite general, and
is applicable to any body at arbitrary temperature. It also takes into
account retardation effects, which become important for large enough
separation between the bodies. A similar approach was used by Polevoi\cite
{seven} but he obtained a nonzero friction force only in the relativistic
limit, in contrast to the present calculations and the earlier calculations
of Persson and Zhang\cite{two}, and Pendry\cite{five}. Polevoi did not give
enough details of his calculation to compare his theory with the present
calculation, but we believe that he overlooked effects related to the change
in the reflectivity of electromagnetic waves from moving bodies, which occur
even in the nonrelativistic limit. In the nonretarded limit and for zero
temperature the present calculation agree with the results of Pendry.
Similarly, in the nonretarded limit and for low sliding velocities, we agree
with the study of Persson and Zhang.

\section{Calculation of the fluctuating electromagnetic field}

We consider two semi-infinite solids having flat parallel surfaces separated
by a distance $d$ and moving with velocity $V$ relative to each other, see
Fig. 1. We introduce the two coordinate systems $K$ and $K^{^{\prime }}$
with coordinate axes $xyz$ and $x^{^{\prime }}y^{^{\prime }}z^{^{\prime }}$
.In the $K$ system body \textbf{1} is at rest while body \textbf{2} is
moving with velocity $V$ along the $x-$ axis ( $xy-$ and $x^{^{\prime
}}y^{^{\prime }}-$ planes are in the surface of body \textbf{1, }$x-$ and $%
x^{^{\prime }}$ axes have the same direction and the $z-$ and $z^{^{\prime
}}-$ axes pointing toward body \textbf{2). }In the $K^{^{\prime }}$ system
body \textbf{2} is at rest while body \textbf{1} is moving with velocity $-V$
along the $x-$ axis. Following to Lifshitz to calculate the fluctuating
field in the interior of the two bodies, we shall use the general theory
which is due to Rytov and is described in detail in his book \cite{nine}.
This method is based on the introduction into the Maxwell equations of a
''random'' field (just as, for example, one introduce a ''random'' force in
the theory of Brownian motion). In the $K$ system in a dielectric,
nonmagnetic body \textbf{1 }for a monochromatic field (time factor $%
e^{-i\omega t}$ ) these equations are 
\begin{eqnarray}
\mathbf{\nabla \times E}_1 &=&i\frac \omega c\mathbf{B}_{1,}  \nonumber \\
\mathbf{\nabla \times B}_1 &=&-i\frac \omega c\varepsilon _1(\omega )\mathbf{%
E}_1\mathbf{-}i\frac \omega c\mathbf{F}_1\mathbf{,}  \label{one}
\end{eqnarray}
where $\mathbf{E}$ and $\mathbf{B}$ are the electric and magnetic fields, $%
\varepsilon _1(\omega )$ is the complex dielectric constant for body $%
\mathbf{1,}$ and $\mathbf{F}$ is the random field. Accordingly to Rytov the
correlation function of the latter, determining the average value of the
product of components of $\mathbf{F}$ at two different points in space, is
given by the formula 
\begin{eqnarray}
&&\left\langle F_i(x,y,z)F_k(x^{^{\prime }},y^{^{\prime }},z^{^{\prime
}})\right\rangle  \nonumber \\
&=&4\hbar \left( \frac 12+n(\omega )\right) \varepsilon ^{^{\prime \,\prime
}}(\omega )\delta _{ik}\delta (x-x^{^{\prime }})\delta (y-y^{^{\prime
}})\delta (z-z^{^{\prime }}),  \label{two} \\
n(\omega ) &=&\frac 1{e^{\hbar \omega /k_BT}-1},  \nonumber
\end{eqnarray}
where $T$ is the temperature and $\varepsilon ^{^{\prime \,\prime }}$ is the
imaginary part of $\varepsilon =\varepsilon ^{^{\prime }}+\varepsilon
^{^{\prime \,\prime }}$. The function $\mathbf{F(}x,y,z\mathbf{)}$\textbf{\ }%
can be represented in the form of a Fourier integral, which can be written
for the half space $z<0$ in the form 
\begin{equation}
\mathbf{F(}x,y,z\mathbf{)=}\int_{-\infty }^{+\infty }\mathbf{g(k)}e^{i%
\mathbf{q\cdot r}}\cos k_zz\,d^3k,  \label{three}
\end{equation}
where a two dimensional vector $\mathbf{q}$ is parallel to the surface, $%
k^2=k_z+q^2,$ and $\mathbf{r}$ is the radius vector in the $x-y$ plane. For
the Fourier components $\mathbf{g(k)}$ , the correlation function
corresponding to the spatial correlation (3) is 
\begin{equation}
\left\langle g_i(\mathbf{k,}\omega \mathbf{)}g_k^{*}(\mathbf{k}^{^{\prime }}%
\mathbf{,}\omega \mathbf{)}\right\rangle =\frac{\hbar \left( \frac
12+n(\omega )\right) \varepsilon ^{^{\prime \,\prime }}(\omega )}{\pi ^3}%
\delta _{ik}\delta (\mathbf{k-k}^{^{\prime }})  \label{four}
\end{equation}
In body \textbf{1 (}$z<0)$ the fields $\mathbf{E}$ and $\mathbf{B}$ can be
written in the form \cite{ten} 
\begin{eqnarray}
\mathbf{E}_1 &=&\int_{-\infty }^{+\infty }\left\{ \mathbf{a}_1(\mathbf{%
k)\cos }k_zz+i\mathbf{b}_1\left( \mathbf{k}\right) \sin k_zz\right\} e^{i%
\mathbf{q\cdot r}}\,d^3k  \nonumber \\
&&+\int_{-\infty }^{+\infty }\mathbf{u}_1(\mathbf{q})e^{i\mathbf{q}\cdot 
\mathbf{r-}is{}_{1\,}z}d^2q,  \label{five}
\end{eqnarray}
\begin{eqnarray}
\mathbf{B}_1 &=&\frac c\omega \int_{-\infty }^{+\infty }\left\{ ([\mathbf{%
q\times a}_1]+k_z[\mathbf{e}_z\times \mathbf{b}_1])\cos k_zz\right. 
\nonumber \\
&&\left. +i([\mathbf{q}\times \mathbf{b}_1]+k_z[\mathbf{e}_z\times \mathbf{a}%
_1])\sin k_zz\right\} e^{i\mathbf{q\cdot r}}d^3k  \nonumber \\
&&+\frac c\omega \int_{-\infty }^{+\infty }\left\{ [\mathbf{q\times u}%
_1]-s_1[\mathbf{e}_z\times \mathbf{u}_1]\right\} e^{i\mathbf{q\cdot r-}%
is_1\,z}d^2\ q,  \label{six}
\end{eqnarray}
where $\mathbf{e}_z$ is a unit vector in the direction of the $z$ axis, and 
\begin{equation}
s_1=\sqrt{\frac{\omega ^2}{c^2}\varepsilon _1-q^2},  \label{seven}
\end{equation}
where the sign of the root is to be chosen so that the imaginary part of $s$
will be positive.

The first ter)s in the expressions (5) and (6) represent a solution of the
inhomogeneous equations (1). Substituting them in the second equation of (1)
and writing $\mathbf{F}$ in the form (3) , one can find the following
relations, expressing $\mathbf{a}_1$ and $\mathbf{b}_1$ in terms of the
Fourier components $\mathbf{g}_1$ of the random field 
\begin{equation}
\mathbf{a}_1=\frac 1{\varepsilon _1(k^2-\omega ^2\varepsilon _1/c^2)}\left[ 
\frac{\omega ^2}{c^2}\varepsilon _1\mathbf{g}_1-\mathbf{q(q\cdot g}%
_1)-k_z^2g_{1z}\mathbf{e}_z\right] ,  \label{eight}
\end{equation}
\begin{equation}
\mathbf{b}_1=-\frac{k_z}{\varepsilon _1(k^2-\omega ^2\varepsilon _1/c^2)}%
\left[ \mathbf{e}_z(\mathbf{q\cdot g}_1\mathbf{)+q}g_{1z}\right] .
\label{nine}
\end{equation}

The second integrals in (5,6) represent the solution of the homogeneous
equations (1) (i.e. the equations with $\mathbf{F}$ omitted), and describe
the plane wave field reflected from the boundary of body. The condition for
transversality of these waves is 
\begin{equation}
\mathbf{u}_1\cdot \mathbf{q-}s_1u_{1z}=0.  \label{ten}
\end{equation}

In the space between bodies(vacuum) $\varepsilon =0,\,\mathbf{F=0}$ and the
field in the $K$ system is given by the general solution of the homogeneous
equations, which can be written in the form 
\begin{equation}
\mathbf{E}_3=\int_{-\infty }^{+\infty }\left\{ \mathbf{v(q,}\omega )e^{ipz}+%
\mathbf{w(q,}\omega )e^{-ipz}\right\} e^{i\mathbf{q\cdot r}}d^2q,
\label{eleven}
\end{equation}
\begin{eqnarray}
\mathbf{B}_3 &=&\frac c\omega \int_{-\infty }^{+\infty }\left\{ \left( [%
\mathbf{q\times v]+p[e}_z\times \mathbf{v]}\right) e^{ipz}\right.  \nonumber
\label{t} \\
&&\left. +\left( [\mathbf{q\times w]-p[e}_z\times \mathbf{w]}\right)
e^{-ipz}\right\} e^{i\mathbf{q\cdot r}}d^2q,  \label{twelve}
\end{eqnarray}
where 
\begin{equation}
p=\sqrt{\frac{\omega ^2}{c^2}-q^2},  \label{thirteen}
\end{equation}
and $\mathbf{v}$ and $\mathbf{w}$ satisfy the transversality conditions 
\begin{equation}
\mathbf{v\cdot q+}pv_z=0,\;\mathbf{w\cdot q-}pw_z=0.  \label{fourteen}
\end{equation}

The boundary conditions on the surfaces of the media are the requirement of
continuity of the tangential components of $\mathbf{E}$ and $\mathbf{B}$ in
the rest frame of respective body. In the $K$ system on the plane $z=0$ for
a given value of $\mathbf{q}$ it is convenient to write the corresponding
equations for components of the fields along the vectors $\mathbf{e}_{%
\mathbf{q}}=\mathbf{q}/q$ and $\mathbf{e}_{\mathbf{n}}=[\mathbf{e}_z\times 
\mathbf{e}_{\mathbf{q}}]$, this gives the following equations 
\begin{eqnarray*}
\int_{-\infty }^{+\infty }a_{1\mathbf{q}}dk_z+u_{1\mathbf{q}} &=&v_{\mathbf{q%
}}+w_{\mathbf{q}}, \\
\int_{-\infty }^{+\infty }a_{1\mathbf{n}}dk_z+u_{1\mathbf{n}} &=&v_{\mathbf{n%
}}+w_{\mathbf{n}},
\end{eqnarray*}
\begin{eqnarray*}
&&\int_{-\infty }^{+\infty }(qa_{1z}-k_zb_{1\mathbf{q}})dk_z+qu_{1z}+s_1u_{1%
\mathbf{q}} \\
&=&q(v_z+w_z)-p(v_{\mathbf{q}}-w_{\mathbf{q}}),
\end{eqnarray*}
\begin{equation}
\ \int_{-\infty }^{+\infty }-k_zb_{1\mathbf{n}}dk_z+s_1u_{1\mathbf{n}}=-p(v_{%
\mathbf{n}}-w_{\mathbf{n}}),  \label{fifteen}
\end{equation}
where $a_{1\mathbf{q}}=\mathbf{e}_{\mathbf{q}}\cdot \mathbf{a}_1,\,a_{1%
\mathbf{n}}=\mathbf{e}_{\mathbf{n}}\cdot \mathbf{a}_1$ and so on. In what
follow we shall need only the field between two media. Using the
transversality conditions (10) and (14) and the expressions (8,9) from the
equations (15) we can obtain the following equations 
\begin{eqnarray}
&&ps_1\int_{-\infty }^\infty \left( qg_{1z}(\mathbf{q,}k_z,\omega )-s_1g_{1%
\mathbf{q}}(\mathbf{q,}k_z,\omega )\right) \frac 1{k_z^2-s_1^2}dk_z 
\nonumber \\
\ &=&-\left( s_1+p\varepsilon _1\right) v_{\mathbf{q}}(\mathbf{q,}\omega
)-\left( p\varepsilon _1-s_1\right) w_{\mathbf{q}}(\mathbf{q,}\omega ),
\label{sixteen}
\end{eqnarray}
. 
\begin{eqnarray}
&&s_1\left( \frac \omega c\right) ^2\int_{-\infty }^\infty \frac{g_{1\mathbf{%
n}}(\mathbf{q,}k_z,\omega )}{k_z^2-s_1^2}dk_z  \nonumber \\
&=&(p+s_1)v_{\mathbf{n}}(\mathbf{q,}\omega )+(s_1-p)w_{\mathbf{n}}(\mathbf{q,%
}\omega ).  \label{seventeen}
\end{eqnarray}

In the $K^{^{\prime }}$ system the Maxwell equations have the same form (1)
and in the second medium ( the half space $z>d$), the field $\mathbf{E}%
_2^{^{\prime }},\,\mathbf{B}_2^{^{\prime }}$ is given by the same formulas
(5-9) with $x-$ coordinate changed to $x^{^{\prime }},$ the index 1 changed
to 2, $\cos k_zz,\,\sin k_zz$ replaced by $\cos k_z(z-d),\,\sin k_z(z-d)$
and change in the sign of $s$ ( the $^{"}$reflected$^{"}$ waves now
propagate along the positive $z$ direction). In the space between media in
the $K^{^{\prime }}$ system the field is given by the same formulas (11-13)
with $x$ changed to $x^{^{\prime }},$ $\mathbf{v,\,w}$ replaced by $\mathbf{v%
}^{^{\prime }}\mathbf{,\,w}^{^{\prime }}$. The relations between the field
in the $K$ and $K^{^{\prime }}$ systems are determined by Lorentz
transformation. Neglecting by the terms of the order $(V/c)^2$ these
relations are given by 
\begin{equation}
\mathbf{v}^{^{\prime }}(\mathbf{q}^{^{\prime }},\omega ^{^{\prime }})=%
\mathbf{v(\mathbf{q}},\omega )+\frac V\omega \left[ \mathbf{e}%
_x\left[ \mathbf{k\times v}(\mathbf{q},\omega )\right] \right] ,
\label{eighteen}
\end{equation}

\begin{equation}
\mathbf{w}^{^{\prime }}(\mathbf{q}^{^{\prime }},\omega ^{^{\prime }})\mathbf{%
=w}(\mathbf{q},\omega )+\frac V\omega \left[ \mathbf{e}_x\left[ \widetilde{%
\mathbf{k}}\times \mathbf{w}(\mathbf{q},\omega )\right] \right] ,
\label{nineteen}
\end{equation}
where $\mathbf{k=(q,}p),\widetilde{\mathbf{k}}=(\mathbf{q,}-p),\omega
^{^{\prime }}=\omega -q_xV,\,\mathbf{q}^{^{\prime }}=\mathbf{q-}\left(
V\omega /c^2\right) \mathbf{e}_x$.

In the $K^{^{\prime }}$ system the boundary conditions at the surface of
body 2 at $z=d$ give the equations 
\begin{eqnarray}
&&ps_2^{-}\int_{-\infty }^\infty \left( q^{^{\prime }}g_{2z}(\mathbf{q}%
^{^{\prime }}\mathbf{,}k_z,\omega ^{^{\prime }})+s_2^{-}q_{2\mathbf{q}%
^{^{\prime }}}(\mathbf{q}^{^{\prime }}\mathbf{,}k_z,\omega ^{^{\prime
}})\right) \frac 1{k_z^2-s_2^{-2}}dk_z  \nonumber \\
&=&(p\varepsilon _2^{-}-s_2^{-})v_{\mathbf{q}^{^{\prime }}}^{^{\prime }}(%
\mathbf{q}^{^{\prime }}\mathbf{,}\omega ^{^{\prime
}})e^{ipd}+(s_2^{-}+p\varepsilon _2^{-})w_{\mathbf{q}^{^{\prime
}}}^{^{\prime }}(\mathbf{q}^{^{\prime }}\mathbf{,}\omega ^{^{\prime
}})e^{-ipd},  \label{twenty}
\end{eqnarray}
\begin{eqnarray}
s_2^{-}\left( \frac{\omega ^{^{\prime }}}c\right) ^2\int_{-\infty }^\infty 
\frac{g_{2\mathbf{n}^{^{\prime }}}(\mathbf{q}^{^{\prime }}\mathbf{,}%
k_z,\omega ^{^{\prime }})}{k_z^2-s_2^{-2}}dk_z &=&(s_2^{-}-p)v_{\mathbf{n}%
^{^{\prime }}}^{^{\prime }}(\mathbf{q}^{^{\prime }}\mathbf{,}\omega
^{^{\prime }})e^{ipd}  \nonumber \\
&&+(s_2^{-}+p)w_{\mathbf{n}^{^{\prime }}}^{^{\prime }}(\mathbf{q}^{^{\prime
}}\mathbf{,}\omega ^{^{\prime }})e^{-ipd},  \label{twentyone}
\end{eqnarray}
where $\varepsilon _2^{-}=\varepsilon _2(\omega -q_xV),\,$%
\begin{equation}
s_2^{-}=\sqrt{(\omega ^{^{\prime }}/c)^2\varepsilon _2(\omega ^{^{\prime
}})-q^{^{\prime }2}}=\sqrt{\frac{(\omega -q_xV)^2}{c^2}(\varepsilon
_2(\omega -q_xV)-1)+p^2,}  \label{twentytwo}
\end{equation}
$p$ is invariant under Lorentz transformation. Now from the equations
(18-19) with accuracy to the terms of the first order in $V/c$ we have 
\begin{equation}
v_{\mathbf{q}^{^{\prime }}}^{^{\prime }}(\mathbf{q}^{^{\prime }}\mathbf{,}%
\omega ^{^{\prime }})=(\mathbf{v}^{^{\prime }}\mathbf{\cdot e}_{\mathbf{q}%
^{^{\prime }}})\approx v_{\mathbf{q}}(\mathbf{q,}\omega )+\frac{q_yp^2V}{%
\omega q^2}v_{\mathbf{n}}(\mathbf{q,}\omega ),  \label{twentythree}
\end{equation}
\begin{equation}
v_{\mathbf{n}^{^{\prime }}}^{^{\prime }}=(\mathbf{v}^{^{\prime }}\cdot 
\mathbf{e}_{\mathbf{n}^{^{\prime }}})\approx \frac{\omega ^{^{\prime }}}%
\omega v_{\mathbf{n}}-\frac{\omega q_yV}{c^2q^2}v_{\mathbf{q}}.
\label{twentyfour}
\end{equation}
The similar equations can be written for $w_{\mathbf{q}^{^{\prime
}}}^{^{\prime }},w_{\mathbf{n}^{^{\prime }}}^{^{\prime }}$ .After
substituting (23) and (24) into the equations (16-17) and (20-21) we get the
system of four equations. These equations can be solved considering the
second terms in the equations (23,24) as a small perturbation. In zero order
we neglect by the second terms . The zero order solution has the form: 
\begin{eqnarray}
v_{\mathbf{q}}^0 &=&\int_{-\infty }^\infty \frac p\Delta \left\{
s_1e^{-ipl}(s_2^{-}+\varepsilon _2^{-}p)\frac{qg_{1z}(\mathbf{q,}k_z,\omega
)-s_1g_{1\mathbf{q}}(\mathbf{q,}k_z,\omega )}{k_z^2-s_1^2}\right.  \nonumber
\\
&&\ \left. +s_2^{-}(\varepsilon _1p-s_1)\frac{q^{^{\prime }}g_{2z}(\mathbf{q}%
^{^{\prime }}\mathbf{,}k_z,\omega ^{^{\prime }})+s_2^{-}g_{2\mathbf{q}%
^{^{\prime }}}(\mathbf{q}^{^{\prime }}\mathbf{,}k_z,\omega ^{^{\prime }})}{%
k_z^2-s_2^{-2}}\right\} dk_z,  \label{twentyfive}
\end{eqnarray}
\begin{eqnarray}
w_{\mathbf{q}}^0 &=&\int_{-\infty }^\infty \frac p\Delta \left\{
-s_1e^{ipd}(\varepsilon _2^{-}p-s_2^{-})\frac{qg_{1z}(\mathbf{q,}k_z,\omega
)-s_1g_{1\mathbf{q}}(\mathbf{q,}k_z,\omega )}{k_z^2-s_1^2}\right.  \nonumber
\\
&&\left. -s_2^{-}(\varepsilon _1p+s_1)\frac{q^{^{\prime }}g_{2z}(\mathbf{q}%
^{^{\prime }}\mathbf{,}k_z,\omega ^{^{\prime }})+s_2^{-}g_{2\mathbf{q}%
^{^{\prime }}}(\mathbf{q}^{^{\prime }}\mathbf{,}k_z,\omega ^{^{\prime }})}{%
k_z^2-s_2^{-2}}\right\} dk_z,  \label{twentysix}
\end{eqnarray}
\begin{eqnarray}
v_{\mathbf{n}}^0 &=&\int_{-\infty }^\infty \frac \omega {c^2\Delta
^{^{\prime }}}\left\{ -\omega s_1e^{-ipd}(s_2^{-}+p)\frac{g_{1\mathbf{n}}(%
\mathbf{q,}k_z,\omega )}{k_z^2-s_1^2}\right.  \nonumber \\
&&\ \left. +\omega ^{^{\prime }}s_2^{-}(s_1-p)\frac{g_{2\mathbf{n}^{^{\prime
}}}(\mathbf{q}^{^{\prime }}\mathbf{,}k_z,\omega ^{^{\prime }})}{%
k_z^2-s_2^{-2}}\right\} dk_z,  \label{twentyseven}
\end{eqnarray}
\begin{eqnarray}
w_{\mathbf{n}}^0 &=&\int_{-\infty }^\infty \frac \omega {c^2\Delta
^{^{\prime }}}\left\{ \omega s_1e^{ipd}(s_2^{-}-p)\frac{g_{1\mathbf{n}}(%
\mathbf{q,}k_z,\omega )}{k_z^2-s_1^2}\right.  \nonumber \\
&&\ \left. -\omega ^{^{\prime }}s_2^{-}(s_1+p)\frac{g_{2\mathbf{n}^{^{\prime
}}}(\mathbf{q}^{^{\prime }}\mathbf{,}k_z,\omega ^{^{\prime }})}{%
k_z^2-s_2^{-2}}\right\} dk_z,  \label{twentyeight}
\end{eqnarray}
\begin{equation}
v_z=-\frac{qv_{\mathbf{q}}}p,\;\;\;w_z=\frac{qw_{\mathbf{q}}}p,
\label{twentynine}
\end{equation}
where we have introduce the notation: 
\[
\Delta =e^{ipd}(\varepsilon _1p-s_1)(\varepsilon
_2^{-}p-s_2^{-})-e^{-ipd}(\varepsilon _1p+s_1)(\varepsilon _2^{-}p+s_2^{-}), 
\]
\[
\Delta ^{^{\prime }}=e^{ipd}(s_1-p)(s_2^{-}-p)-e^{-ipd}(p+s_1)(p+s_2^{-}). 
\]
The first order solution has the form: 
\begin{eqnarray}
v_{\mathbf{q}}^1 &=&\frac{(p\varepsilon _1-s_1)\Lambda }\Delta ,\;w_{\mathbf{%
q}}^1=-\frac{(p\varepsilon _1+s_1)\Lambda }\Delta ,  \label{thirty} \\
v_{\mathbf{n}}^1 &=&\frac{(s_1-p)\Lambda ^{^{\prime }}}{\Delta ^{^{\prime }}}%
,\;w_{\mathbf{n}}^1=-\frac{(p+s_1)\Lambda ^{^{\prime }}}{\Delta ^{^{\prime }}%
},  \label{thirtyone}
\end{eqnarray}
where 
\begin{eqnarray*}
\Lambda &=&-\frac{q_yp^2V}{\omega q^2}\left[ (p\varepsilon _2^{-}-s_2^{-})v_{%
\mathbf{n}}^0e^{ipd}+(p\varepsilon _2^{-}+s_2^{-})w_{\mathbf{n}%
}^0e^{-ipd}\right] , \\
\Lambda ^{^{\prime }} &=&\frac{\omega ^{^{\prime }}q_yV}{c^2q}\left[
(s_2^{-}-p)v_{\mathbf{q}}^0e^{ipd}+(s_2^{-}+p)w_{\mathbf{q}%
}^0e^{-ipd}\right] .
\end{eqnarray*}

\section{Calculation of the force of friction}

The frictional stresses $\sigma $ and $-\sigma $ which act on the surfaces
of the two bodies can be obtained from the $xz$-component of the Maxwell
stress tensor $\sigma _{ij}$, evaluated at $z=0$: 
\begin{equation}
\sigma =\frac 1{8\pi }\int_{-\infty }^{+\infty }d\omega [\langle
E_{3z}E_{3x}^{*}+\langle E_{3z}^{*}E_{3x}\rangle +\left\langle
B_{3z}B_{3x}^{*}\right\rangle +\langle B_{3z}^{*}B_{3x}\rangle ]_{z=0}
\label{thirtytwo}
\end{equation}
Here $\langle ..\rangle $ denote a statistical average over the random
field. The averaging is carried out with the aid of equation(4). Note that
the components of the random field $\mathbf{g}_1$ and $\mathbf{g}_2$
referring to different media are statistically independent, so the average
of their product gives zero. Writing the squares of the integrals (11-12) in
the usual way as double integrals, and carrying out one integration over the 
$\delta -$ function, we obtain 
\begin{eqnarray}
\sigma  &=&\frac 1{8\pi }\int d\omega d^2q[\langle E_{3z}(\mathbf{q},\omega
)E_{3x}^{*}(\mathbf{q},\omega )\rangle +\langle E_{3z}^{*}\mathbf{q},\omega
)E_{3x}(\mathbf{q},\omega )\rangle   \nonumber \\
&&\ +\langle B_{3z}(\mathbf{q},\omega )B_{3x}^{*}(\mathbf{q},\omega
)+\langle B_{3z}^{*}\mathbf{(q},\omega )B_{3x}(\mathbf{q},\omega )\rangle
]_{z=0},  \label{thirtythree}
\end{eqnarray}
where one must substitute in place of $\mathbf{E}_{3}$ and $\mathbf{B%
}_3$ ,the expressions in the integrands (11-12) determined by the formulas
(25-31), and the average product $\left\langle g_i(\mathbf{k,}\omega
)g_k^{*}(\mathbf{k,}\omega )\right\rangle $ is to be taken as $(1/2+n(\omega
))\varepsilon ^{^{\prime \prime }}(\omega )\delta _{ik}/\pi ^3.$ For a given
value of $\mathbf{q}$ it is convenient to express the components $E_x$ and $%
B_x$ in terms of the components along the two vectors $\mathbf{e}_{\mathbf{q}%
}$ and $\mathbf{e}_{\mathbf{n}}$ 
\begin{eqnarray*}
E_x &=&(q_x/q)E_{\mathbf{q}}-(q_y/q)E_{\mathbf{n}}, \\
B_x &=&(q_x/q)B_{\mathbf{q}}-(q_y/q)B_{\mathbf{n}}.
\end{eqnarray*}

Thus we can write 
\begin{eqnarray}
\sigma &=&\frac 1{8\pi }\int d\omega d^2q\left\{ \frac{q_x}q[\langle E_z(%
\mathbf{q},\omega )E_{\mathbf{q}}^{*}(\mathbf{q},\omega )\rangle +\langle
E_z^{*}q,\omega )E_{\mathbf{q}}(\mathbf{q},\omega )\rangle \right.  \nonumber
\\
&&+\langle B_z(\mathbf{q},\omega )B_{\mathbf{q}}^{*}(\mathbf{q},\omega
)\rangle +\langle B_z^{*}(\mathbf{q},\omega )B_{\mathbf{q}}(\mathbf{q}%
,\omega )\rangle ]_{z=0}  \nonumber \\
&&-\frac{q_y}q\left[ \langle E_z(\mathbf{q},\omega )E_{\mathbf{n}}^{*}(%
\mathbf{q},\omega )\rangle +\langle E_z^{*}q,\omega )E_{\mathbf{n}}(\mathbf{q%
},\omega )\rangle \right.  \nonumber \\
&&\left. +\langle B_z(\mathbf{q},\omega )B_{\mathbf{n}}^{*}(\mathbf{q}%
,\omega )\rangle +\langle B_z^{*}\mathbf{(q},\omega )B_{\mathbf{n}}(\mathbf{q%
},\omega )\rangle ]_{z=0}\right\} .  \label{thirtyfour}
\end{eqnarray}

Accordingly t) eq.(11-12)

\begin{eqnarray}
E_z &=&(v_z+w_z)=(q/p)(w_{\mathbf{q}}-v_{\mathbf{q}})=(qp^{*}/\mid p\mid
^2)(w_{\mathbf{q}}-v_{\mathbf{q}}),  \nonumber \\
E_{\mathbf{q}} &=&v_{\mathbf{q}}+w_{\mathbf{q}},  \nonumber \\
E_{\mathbf{n}} &=&v_{\mathbf{n}}+w_{\mathbf{n}},  \nonumber \\
B_z &=&(cq/\omega )(v_{\mathbf{n}}+w_{\mathbf{n}}),  \nonumber \\
B_{\mathbf{q}} &=&(cp/\omega )(w_{\mathbf{n}}-v_{\mathbf{n}}),  \nonumber \\
B_{\mathbf{n}} &=&(\omega p^{*}/c\left| p\right| ^2)(v_{\mathbf{q}}-w_{%
\mathbf{q}}).  \label{thirtyfive}
\end{eqnarray}

After substituting these expressions into formula (34) one can see that the
second term with $q_y$ is identically equal to zero . From equations (25-31)
it follows that the zero and first order solutions are statistically
independent, then, neglecting by the terms of the order $(v/c)^2$ ,from
(34-35) we obtain 
\begin{eqnarray*}
\sigma &=&\frac 1{4\pi }\int_0^{+\infty }d\omega \int d^2qq_x\left( \left[
\frac 1{\left| p\right| ^2}(p+p^{*})(\left\langle \mid w_{\mathbf{q}}^0\mid
^2\right\rangle -\left\langle \mid v_{\mathbf{q}}^0\mid ^2\right\rangle
)\right. \right. \\
&&\left. +(p-p^{*})\left\langle (v_{\mathbf{q}}^0w_{\mathbf{q}}^{0*}-v_{%
\mathbf{q}}^{0*}w_{\mathbf{q}}^0)\right\rangle \right]
\end{eqnarray*}
\begin{equation}
\left. +\left( \frac c\omega \right) ^2\left[ (p+p^{*})(\left\langle \mid w_{%
\mathbf{n}}^0\mid ^2\right\rangle -\left\langle \mid v_{\mathbf{n}}^0\mid
^2\right\rangle )-(p-p^{*})\left\langle (v_{\mathbf{n}}^0w_{\mathbf{n}%
}^{0*}-v_{\mathbf{n}}^{0*}w_{\mathbf{n}}^0)\right\rangle \right] \right) ,
\label{thirtysix}
\end{equation}
where we change the integration over $d\omega $ between the limits $-\infty $
and $+\infty $ on the integration only over positive values of $\omega $
what gives the extra factor two.

Taking into account that $p=p^{*}$ for $q<\omega /c$ , $p=-p^{*}$ for $%
q>\omega /c,$ and carrying out the integration over $dk_z$ with the help of
the formula 
\[
\int_{-\infty }^\infty \frac{dk_z}{\left| k_z^2-s^2\right| ^2}=\frac{i\pi }{%
\left| s\right| ^2(s-s^{*})}, 
\]
after substituting (25-31) into (36) we obtain 
\[
\sigma =\frac \hbar {8\pi ^3}\int_0^\infty d\omega \int_{q<\omega /c}d^2qq_x 
\]
\[
\times \left\{ \frac{(1-\mid R_{1p}\mid ^2)(1-\mid R_{2p}^{-}\mid ^2)}{\mid
1-e^{2ipd}R_{1p}R_{2p}^{-}\mid ^2}\left( n(\omega -q_xV)-n(\omega )\right)
+[R_p\rightarrow R_s]\right\} 
\]
\[
+\frac \hbar {2\pi ^3}\int_0^\infty d\omega \int_{q>\omega
/c}d^2qq_xe^{-2\mid p\mid d} 
\]
\begin{equation}
\times \left\{ \frac{\mathrm{Im}R_{1p}\mathrm{Im}R_{2p}^{-}}{\mid
1-e^{-2\mid p\mid d}R_{1p}R_{2p}^{-}\mid ^2}\left( n(\omega -q_xV)-n(\omega
)\right) +[R_p\rightarrow R_s]\right\}  \label{thirtyseven}
\end{equation}
where 
\[
R_{ip}=\frac{\varepsilon _ip-s_i}{\varepsilon _ip+s_i},\ \quad R_{is}=\frac{%
s_i-p}{s_i+p}, 
\]
\[
R_{ip}^{\pm }=\frac{\varepsilon _i^{\pm }p-s_i^{\pm }}{\varepsilon _i^{\pm
}p+s_i^{\pm }},\ \quad R_{is}^{\pm }=\frac{s_i^{\pm }-p}{s_i^{\pm }+p}, 
\]
$\varepsilon _i^{\pm }(\omega )=\varepsilon _i(\omega \pm q_xV)$ and $%
s_i^{\pm }(\omega )=s_i(\omega \pm q_xV)$, $i=1,2$. Note that $R_{ip}$ and $%
R_{is}$ are the electromagnetic reflection factors for $p$-polarized and $s$%
-polarized light, respectively. ($p$-polarized light has the electric field
vector in the plane of incidence while the electric field vector is
perpendicular to this plane for $s$-polarized light.)The first term in (37)
is the contribution to the friction force from the propagating (radiating)
electromagnetic field, i.e., the black body radiation. This term includes
only the thermal radiation and is equal to zero at $T=0.$. The second term
is derived from the evanescent field, i.e., from the component of the
electromagnetic field which decay exponentially with the distances away from
the surfaces of the bodies. This term does not vanish even at $T=0$K because
of quantum fluctuations in the charge density in the solids.

\section{Some limiting cases}

Let us first consider distances $d<<c/\omega _p$, where $\omega _p$ is the
plasma frequency of the metals. For typical metals, $c/\omega _p\approx 200\ 
\mathrm{\AA }$. In this case the main contribution comes from $q>>\omega
_p/c $, and we have $s_1\approx s_2\approx p\approx iq$, $R_s\approx 0$ and 
\[
R_p\approx \frac{\varepsilon -1}{\varepsilon +1} 
\]
In this approximation the integration over $d^2q$ can be extended on the all 
$q-$ plane. Using these approximations, the second term in (11) can be
written as 
\[
\sigma =\frac \hbar {4\pi ^3}\int_0^\infty d\omega \int d^2qq_xe^{-2\mid
p\mid d} 
\]
\[
\times \left\{ \left( \frac{\mathrm{Im}R_{1p}\mathrm{Im}R_{2p}^{-}}{\mid
1-e^{-2\mid p\mid d}R_{1p}R_{2p}^{-}\mid ^2}+(1\leftrightarrow 2)\right)
\left( n(\omega -q_xV)-n(\omega )\right) +[R_p\rightarrow R_s]\right\} 
\]
\[
=\frac \hbar {2\pi ^3}\int_{-\infty }^\infty dq_y\int_0^\infty
dq_xq_xe^{-2qd}\left\{ \int_0^\infty d\omega [n(\omega )-n(\omega
+q_xv)]\right. 
\]
\[
\times \left( \frac{\mathrm{Im}R_{1p}^{+}\mathrm{Im}R_{2p}}{\mid
1-e^{-2\mid p\mid d}R_{1p}^{+}R_{2p}\mid ^2}+\left( 1\leftrightarrow
2\right) \right) 
\]
\begin{equation}
\left. -\int_0^{q_xv}d\omega [n(\omega )+1/2]\left( \frac{\mathrm{Im}%
R_{1p}^{-}\mathrm{Im}R_{2p}}{\mid 1-e^{-2qd}R_{1p}^{-}R_{2p}\mid ^2}%
+(1\leftrightarrow 2)\right) \right\}  \label{thirtyeight}
\end{equation}
where we have used the relation $n(-\omega )=-n(\omega )-1.$ At zero
temperature the Bose-Einstein factor $n(\omega )=0$ and only the second term
in (38) will contribute to the sliding friction; in this limit our
expression for the friction force for two identical solids is in agreement
with Pendry [his result is, however, not symmetric with respect to $%
(1\leftrightarrow 2)$ which must be the case because of symmetry in the
nonretarded limit]. In appendix A we show that the zero-temperature result
can be generalized to include nonlocal optics effects, by replacing the
reflection factor $R_p(\omega )$ in (38) by the surface response function $%
g(q,\omega )$. Next, let us consider the limiting case of low sliding
velocity or high temperature namely, $V<<cd/d_W$, where $d_W=c\hbar /k_BT$
is the Wien length (typically $d_W\approx 10^5\AA $). In this case 
\[
n(\omega )-n(\omega +q_xV)\approx -q_xV\frac{dn}{d\omega }=\frac{e^{\hbar
\omega /k_BT}}{\left( e^{\hbar \omega /k_BT}-1\right) ^2}\frac{\hbar q_xV}{%
k_BT} 
\]
and in the second term in (38) we can put 
\[
n(\omega )\approx k_BT/\hbar \omega 
\]
Substituting these results in (38) gives 
\[
\sigma =\frac{\hbar V}{2\pi ^2}\int_0^\infty dqq^3e^{-2qd}\int_0^\infty
d\omega \left( -\frac{dn}{d\omega }\right) \frac{\mathrm{Im}R_{1p}\mathrm{%
Im}R_{2p}}{\mid 1-e^{-2qd}R_{1p}R_{2p}\mid ^2} 
\]
\begin{equation}
+\frac 2{\pi ^3}k_BT\int_{-\infty }^\infty dq_y\int_0^\infty
dq_xq_xe^{-2qd}\int_0^{q_xV}\frac{d\omega }\omega \frac{\mathrm{Im}R_{1p}%
\mathrm{Im}R_{2p}}{\mid 1-e^{-2qd}R_{1p}R_{2p}\mid ^2}  \label{thirtynine}
\end{equation}
The second term in this expression is proportional to $\sim V^2$ as $%
V\rightarrow 0$ (see below) and can be neglected in the limit of small $V$.
The first term is $\sim V$ and is in agreement with the result obtained by
Persson and Zhang if one assumes local optics (which implies replacing $%
g(q,\omega )\rightarrow R_p(\omega )$ in \cite{two}). For free-electron like
metals the local optics is accurate if $d>>l$, where $l$ is the electron
mean free path in the metal. If this condition is not satisfied, the general
formula of Persson and Zhang must be used.

Let us consider two identical metals described by the dielectric function 
\begin{equation}
\varepsilon (\omega )=1-\frac{\omega _p^2}{\omega (\omega +i/\tau )},
\label{fourty}
\end{equation}
where $\tau $ is the relaxation time and $\omega _p$ the plasma frequency.
Thus, for small frequencies 
\[
\mathrm{Im}R_p\approx \frac{2\omega }{\omega _p^2\tau },\ \ \ \ \mathrm{Re}%
R_p\approx 1. 
\]
and if we neglect the imaginary part of $R_p$ in the denominator of the
integrand in (13) we obtain 
\begin{equation}
\sigma =\xi \left( \frac{k_BT}{\hbar \omega _p}\right) ^2\frac 1{(\omega
_p\tau )^2}\frac{\hbar V}{d^4}+\frac{2\pi }{45}\frac{k_BT}{\hbar \omega _p}%
\frac 1{(\omega _p\tau )^2}\frac{\hbar V^2}{\omega _pd^5},  \label{fourtyone}
\end{equation}
where 
\[
\xi =\frac 18\int_0^\infty \frac{dxx^2}{e^x-1}\approx 0.5986. 
\]
In deriving (41) we have used the following standard integrals 
\[
\int_0^\infty \frac{dxx}{e^x-1}=\frac{\pi ^2}6,\ \ \ \ \int_0^\infty \frac{%
dxx^3}{e^x-1}=\frac{\pi ^4}{15}. 
\]
The ratio between the second and first term in (41) equals $\approx
(V/c)(d_W/d)$, and in deriving (41) we have assumed that this quantity is
much smaller than unity. As an example, if $d=10\mathrm{\AA }$ and $V=1%
\mathrm{m/s}$ then for typical metals at room temperature ($k_BT\approx 0.025%
\mathrm{\ eV}$, $\omega _p\tau \approx 100$, $\hbar \omega _p\approx 10\ 
\mathrm{eV}$) the first and the second terms in (41) give $\sigma \approx
10^{-8}$ and $\approx 10^{-13}\mathrm{N/m}^2$, respectively.

On the other hand, if $(V/c)(d_W/d)>>1$ we get 
\begin{equation}
\sigma =\frac \xi 2\left( \frac{k_BT}{\hbar \omega _p}\right) ^2\frac
1{(\omega _p\tau )^2}\frac{\hbar V}{d^4}+\zeta \frac 1{(\omega _p\tau )^2}%
\frac{\hbar V}{d^4}\left( \frac V{d\omega _p}\right) ^2,  \label{fourtytwo}
\end{equation}
where 
\[
\zeta =\frac 5{2^9\pi ^2}\int_0^\infty \frac{dxx^4}{e^x-1}=0.024610. 
\]
The ratio of the second and first term in (42) equals $\sim
0.1(V/c)^2(d_W/d)^2$. It is clear that at low temperature or high
velocities, the second term in (42) will dominate.

Next, let us consider the sliding friction to linear order in the sliding
velocity when $c/\omega _p<<d<<d_W$. There will be two contributions
associated with $R_p$ and $R_s$. As shown in Appendix B, the contribution
from $R_p$ is 
\begin{equation}
\sigma _p\approx \frac{3\xi }{\pi ^2}\frac{\hbar V}{d^4}\left( \frac
d{d_W}\right) ^2\frac{k_BT}{\hbar \omega _p}\frac 1{\omega _p\tau }\left(
1+\frac 1e+\ln \frac{d_W}d\right) .  \label{fourtythree}
\end{equation}
The contribution $\sigma _s$ from the term involving $R_s$ is given by (see
Appendix B)

\begin{equation}
\sigma _s\approx C\omega _p\tau \frac Vc\frac{\hbar \omega _p}{d^2d_W},
\label{fourtyfour}
\end{equation}
where $C\approx 0.394$. Comparing (43) with (44) we obtain 
\[
\sigma _s/\sigma _p\approx (\omega _p\tau )^2(\hbar \omega _p/k_BT)^2. 
\]
For typical metals at room temperature, $\hbar \omega _p/k_BT\sim 10^3$ and $%
\omega _p\tau \sim 100$ so that $\sigma _s/\sigma _p\sim 10^{10}$, i.e., the
main contribution comes from the term involving $R_s$. As an illustration,
if $d=10^4\mathrm{\AA }$ and $V=1\mathrm{m/s}$ then for metals at room
temperature, characterized by the same parameter values as used above, one
get $\sigma _s\approx 10^{-8}\mathrm{N/m}^2$.

Now, let us consider the radiative contribution to the friction force, which
is given by the first term in (38). In linear order in the sliding velocity
we get 
\[
\sigma _{rad}=\frac{\hbar V}{8\pi ^3}\int_0^\infty d\omega \int_{q<\omega
/c}d^2qq_x^2 
\]
\begin{equation}
\times \left\{ \frac{(1-\mid R_{1p}\mid ^2)(1-\mid R_{2p}\mid ^2)}{\mid
1-e^{2ipd}R_{1p}R_{2p}\mid ^2}\left( -\frac{\partial n(\omega )}{\partial
\omega }\right) +[R_p\rightarrow R_s]\right\} .  \label{fourtyfive}
\end{equation}
For separations $d$ much smaller than the Wien wavelength $d_W=c\hbar /k_BT$
we can put $\exp (ipd)\approx 1$. In this case and for the small
frequencies, when $\omega \leq k_BT/\hbar <<1/\tau ,$we get for identical
metals described by the dielectric function (40) 
\begin{eqnarray}
\frac{\left( 1-\mid R_p\mid ^2\right) ^2}{\mid 1-R_p^2\mid ^2} &=&\frac 12+%
\frac{(\varepsilon ^{*}s)^2+(\varepsilon s^{*})^2}{4\mid \epsilon s\mid ^2}%
\approx \frac 12+\frac{\varepsilon ^{*}+\varepsilon }{4\mid \varepsilon \mid 
}  \nonumber  \label{fourtysix} \\
&\approx &\frac 12\left( 1+\frac{\omega \tau }2\right) \approx \frac 12.
\label{fourtysix}
\end{eqnarray}
and the same result is obtained when $R_p$ in (46) is replaced by $R_s.$ The
final result for the radiative friction force has the form . 
\begin{equation}
\sigma _{rad}=\frac{\hbar V}{8\pi ^2c^4}\int_0^\infty d\omega \omega
^3n(\omega )=\frac{\pi ^2}{120}\hbar V\left( \frac{k_BT}{\hbar c}\right) ^4.
\label{fourtyseven}
\end{equation}
Note that the radiative stress does not depend on the separation and is
proportional to $T^4$.The latter result is, of course, only valid as long as 
$d$ is small compared with the lateral extend (or linear size) $L$ of the
bodies. When $d$ becomes comparable with or larger than $L$, the friction
force between the two bodies will decrease monotonically with increasing $d$%
. At room temperature and at the sliding velocity $V=1m/s$ one get $\sigma
_{rad}\approx 10^{-15}\mathrm{N/m}^2$. The ratio of this contribution to $%
\sigma _s$ from (44) is 
\[
\frac{\sigma _{rad}}{\sigma _s}=0.1\left( \frac d{d_W}\right) ^2\frac{k_BT}{%
\hbar \omega _p}\frac 1{\omega _p\tau }\sim 10^{-6}\left( \frac
d{d_W}\right) ^2. 
\]
Thus for $d\sim d_W$ (which is of order $\sim 10^5\,$ for typical metals at
room temperature) the nonradiative part dominate over the radiative
contribution by a factor $\sim 10^6\mathrm{\,}$. However, for large enough
distances the radiative part dominate as this contribution is finite for
arbitrary separations $d$.

\section{Summary and conclusion}

We have calculated the friction force between two arbitrary bodies with flat
surfaces separated by a vacuum slab of thickness $d$, and moving with a
relative velocity $V$. The separation $d$ is assumed to be so large that the
only interaction between the bodies is via the electromagnetic field
associated with \textit{thermal }or \textit{quantum} fluctuations in the
solids. A general formula for the friction force has been obtained, which is
valid for arbitrary velocity $V$, separation $d$ and temperature $T$, and
applicable to any bodies. At low sliding velocity only thermal fluctuations
give a contribution to the friction force, linearly proportional to the
velocity $V$. Quantum fluctuations give a nonlinear (in $V$) contribution to
the friction force and is usually negligible compared with the thermal
contribution. [There is also a contribution from quantum fluctuations which
is proportional to $V$, resulting from higher order electron-photon
processes than considered in our work. However, this contribution decays as $%
\sim exp(-2Gd)$ (where $G=2\pi /a$ is the smallest reciprocal lattice
vector) and is negligible small already for $d=10\,\mathrm{\,\AA }$. See 
\cite{eleven,two}.] We have studied the detailed distance dependence of the
friction force from short distances, where retardation effects can be
neglected, to large distances where retardation effects and black body
radiation are important. In most practical cases, involving sliding of a
block on a substrate, the van der Waals friction makes a negligible
contribution to the friction force (the main part of the friction arises
from the regions of real contact between the solids). However, in some
special cases the van der Waals friction is very important\cite{two}. For
example, quantum fluctuations contribute in an important manner to the
friction force acting on thin physisorbed layers of atoms sliding on
metallic surfaces \cite{twelve} [In this case there is an overlap of the
wavefunctions of the sliding layer and those of the metal, which result in
second contribution to the friction force, derived from the repulsive
``contact interaction'' (Pauli repulsion) between the sliding layer and the
substrate]. In addition, the contribution from thermal fluctuations gives
the dominating drag force in some experiments involving parallel 2D-electron
systems \cite{thirteen}

\textbf{Acknowledgments}

A.I. Volokitin acknowledges financial support from DFG, the Russian
Foundation of Basic Research (Project No 96-02-16112) and State Science
scholarship for scientist of Russia. B.N.J. Persson acknowledges financial
support from BMBF (Deutsch-Israelische Project-Kooperation ``Neue Strategien
in der Tribologie: von Nano- bis zu Mesoskalen'').

\appendix 

In this appendix we derive an expression for the (nonlinear) sliding
friction using a nonlocal optics description of the metals. For simplicity,
we focus on zero temperature $T=0$K and assume that $d$ is so short that
retardation effects can be neglected (only in this ``short-distance'' region
will nonlocal effects be important). The calculation is based on the
formalism developed in \cite{fourteen,two}. Let us first define the linear
response function $g(q,\omega )$ which is needed below. Assume that a
semi-infinite metal occupy the half space $z\le 0$. A charge distribution in
the half space $z>d$ gives rise to an (external) potential which must
satisfy Laplace equation for $z<d$ and which therefore can be written as a
sum of evanescent plane waves of the form 
\[
\phi _{ext}=\phi _0e^{qz}e^{i\mathbf{q\cdot x}-i\omega t} 
\]
where $\mathbf{q}=(q_x,q_y)$ is a 2D-wavevector. This potential will induce
a charge distribution in the solid (occupying $z<0$) which in turn gives
rise to an electric potential which must satisfy the Laplace equation for $%
z>0$, and which therefore can be expanded into evanescent plane waves which
decay with increasing $z>0$. Thus the total potential for $0<z<d$ can be
expanded in functions of the form 
\[
\phi _{ext}=\phi _0\left( e^{qz}-ge^{-qz}\right) e^{i\mathbf{q\cdot x}%
-i\omega t} 
\]
where the reflection factor $g=g(q,\omega )$. For the present purposes, we
can treat the low-energy electron-hole pair excitations in the metals as
bosons. As shown in Ref. \cite{fourteen}, the Hamiltonian for the total
system can be written as 
\begin{eqnarray}
H &=&\sum_{\mathbf{q}\alpha _1}\hbar \omega _{\mathbf{q}\alpha _1}b_{\mathbf{%
q}\alpha _1}^{+}b_{\mathbf{q}\alpha _1}+\sum_{\mathbf{q}\alpha _2}\hbar
\omega _{\mathbf{q}\alpha _2}b_{\mathbf{q}\alpha _2}^{+}b_{\mathbf{q}\alpha
_2}+\hbar \omega b^{+}b  \nonumber \\
&&\ \ +\sum_{\mathbf{q}\alpha _1n}C_{q\alpha _1}e^{-qz_n}\left( b_{\mathbf{q}%
\alpha _1}e^{i\mathbf{q}\cdot (\mathbf{x}_n+\mathbf{V}t)}+h.c.\right) . 
 \label{Aone}
\end{eqnarray}
Here $\omega _{\mathbf{q}\alpha _1}$, $b_{\mathbf{q\alpha _1}}^{+}$ and $b_{%
\mathbf{q}\alpha _1}$ are the angular frequency and creation and
annihilation operators for the bosons (of solid $\mathbf{1}$) with the
quantum numbers $(\mathbf{q,}\alpha _1\mathbf{)}$, and $C_{\mathbf{q}\alpha
_1}$ parameters determining the coupling between the boson excitations in
solid $\mathbf{1}$ with the electrons in solid $\mathbf{2}$. Similarly, $b_{%
\mathbf{q}\alpha _2}^{+}$ and $b_{\mathbf{q}\alpha _2}$ are creation and
annihilation operators for bosons in solid $\mathbf{2}$, and $(\mathbf{x}_n%
\mathbf{,z}_n\mathbf{)}$ is the position operator of electron $n$ in solid $%
\mathbf{2}$, which in principle could be expressed in terms of the operators 
$b_{\mathbf{q}\alpha _2}^{+}$ and $b_{\mathbf{q}\alpha _2}$, but for the
present purpose this is not necessary. As shown in \cite{fourteen}, $C_{%
\mathbf{q\alpha _1}}$ can be related to $\mathrm{Im}g_1(q,\omega )$ via 
\begin{equation}
\sum_{\alpha _1}\mid C_{\mathbf{q}\alpha _1}\mid ^2\delta (\omega -\omega
_{q\alpha _1})=\frac{2e^2\hbar }{Aq}\mathrm{Im}g_1(q,\omega )  
\label{Atwo}
\end{equation}
We can write the interaction Hamiltonian between solid $\mathbf{1}$ and $%
\mathbf{2}$ as 
\[
H^{^{\prime }}=\sum_{\mathbf{q}}\left( \hat V_{\mathbf{q}}e^{i\mathbf{q\cdot
V}t}+h.c.\right) 
\]
Using time-dependent perturbation theory (with $H^{^{\prime }}$ as the
perturbation) we can calculate the energy transfer from the translational
motion (kinetic energy) to internal excitations in the solids (boson
excitations $\omega _{\mathbf{q}\alpha _1}$ and $\omega _{\mathbf{q}\alpha
_2}$ in solid $\mathbf{1}$ and $\mathbf{2}$, respectively): 
\begin{eqnarray}
P &=&\frac{2\pi }{\hbar ^2}\sum_{\mathbf{q\alpha _1\alpha _2}}\hbar \omega _{%
\mathbf{q}}\delta (\omega _{\mathbf{q}}-\omega _{\mathbf{q}\alpha _2}-\omega
_{\mathbf{q}\alpha _1})\mid C_{q\alpha _1}\mid ^2  \nonumber \\
\times e^{-2qd} &\mid &\langle n_{\mathbf{q}\alpha _1}=1,n_{\mathbf{q}\alpha
_2}=1\mid \sum_ne^{-q(z_n-d)}e^{-i\mathbf{q\cdot x_n}}b_{\mathbf{q}\alpha
_1}^{+}\mid 0,0\rangle \mid ^2    \label{Athree}
\end{eqnarray}
where $\omega _{\mathbf{q}}=\mid \mathbf{q\cdot V\mid }$. To simplify (A3),
let us write 
\begin{equation}
\delta (\omega _{\mathbf{q}}+\omega _{\mathbf{q}\alpha _2}-\omega _{\mathbf{q%
}\alpha _1})=\int d\omega ^{\prime }\delta (\omega ^{\prime }-\omega _{%
\mathbf{q\alpha _1}})\delta (\omega _{\mathbf{q}}-\omega ^{\prime }-\omega _{%
\mathbf{q\alpha _2}})  \label{afour}
\end{equation}
Substituting (A4) in (A3) and using (A2) gives 
\begin{equation}
P=\frac{4\pi e^2}A\sum_{\mathbf{q}}\frac{\omega _{\mathbf{q}}}qe^{-2qd}\int
d\omega ^{\prime }\mathrm{Im}g_1(q,\omega ^{\prime })M_q(\omega _{%
\mathbf{q}}-\omega ^{\prime })    \label{Afive}
\end{equation}
where 
\[
M_q(\omega )=\sum_{\alpha _2}\delta (\omega -\omega _{\mathbf{q}\alpha
_2})\mid \langle n_{\mathbf{q}\alpha _2}=1\mid \sum_ne^{-q(z_n-d)}e^{-i%
\mathbf{q\cdot x_n}}\mid 0\rangle \mid ^2 
\]
But it has been shown elsewhere that \cite{fifteen} 
\[
\frac{A\hbar q}{2\pi ^2e^2}\mathrm{Im}g_2(q,\omega )=\sum_{\alpha _2}\delta
(\omega -\omega _{\mathbf{q}\alpha _2})\mid \langle n_{\mathbf{q}\alpha
_2}=1\mid \sum_ne^{-q(z_n-d)}e^{-i\mathbf{q\cdot x_n}}\mid 0\rangle \mid ^2 
\]
so that 
\begin{equation}
M_q(\omega )=\frac{A\hbar q}{2\pi ^2e^2}\mathrm{\mathrm{Im}}g_2(q,\omega ) 
  \label{Asix}
\end{equation}
Substituting this result in (A5) gives 
\begin{equation}
P=\frac{2\hbar }\pi \sum_{\mathbf{q}}\omega _{\mathbf{q}}e^{-2qd}\int
d\omega ^{\prime }\mathrm{Im}g_1(q,\omega ^{\prime })\mathrm{Im}%
g_2(q,\omega _{\mathbf{q}}-\omega ^{\prime })   \label{Aseven}
\end{equation}
Finally replacing 
\[
\sum_{\mathbf{q}}\rightarrow \frac A{4\pi ^2}\int d^2q 
\]
and using the relation $P\ =\sigma AV$\textrm{\ }between the power P and the
shear stress $\sigma $ gives 
\begin{equation}
\sigma =\frac \hbar {2\pi ^3}\int d^2q\mid q_x\mid e^{-2qd}\int_0^{\mid
q_x\mid V}d\omega ^{\prime }\mathrm{Im}g_1(q,\omega ^{^{\prime }})^{\prime }%
\mathrm{Im}g_2(q,\mid q_x\mid V-\omega ^{^{\prime }})  
\end{equation}
where we have used that $\mathrm{Im}g(q,\omega )=0$ for $\omega <0.$The
coupling $H^{^{\prime }}$ does not only give rise to real excitations but
also to screening (image charge effects). To take these into account one
must go to higher order in perturbation theory. Following Ref.\cite{two}
this gives a modification of (A8) 
\[
\sigma =\frac \hbar {2\pi ^3}\int d^2q\frac{\mid q_x\mid e^{-2qd}}{\mid
1-g_1(q,0)g_2(q,0)e^{-2qd}\mid ^2} 
\]
\begin{equation}
\times \int_0^{\mid q_x\mid V}d\omega ^{\prime }\mathrm{Im}g_1(q,\omega
^{\prime })\mathrm{Im}g_2(q,\mid q_x\mid V-\omega ^{\prime })  
\label{Anine}
\end{equation}
where we have assumed that the small frequencies involved in the real
excitations are screened in an adiabatic manner so that $g_1$ and $g_2$ can
be evaluated at zero frequency in the screening factor. In the case of local
optics, this expression for $\sigma $ agree with the last term in (38)
evaluated at zero temperature. At finite temperature ($T>0$K) an extra
factor of $[1+2n(\omega ^{^{\prime }})]$ must be inserted in the frequency
integral in (A9) to take into account the enhanced probability for
excitation of bosons at finite temperature. For $T>0$K one must, in addition
to the process considered above, also include scattering processes where a
thermally excited boson is annihilate either in solid $\mathbf{1}$ or in
solid\textbf{\ }$\mathbf{2}$, namely $(n_{\mathbf{q}\alpha _1}=0,n_{\mathbf{q%
}\alpha _2}=1)\rightarrow (1,0)$ and $(1,0)\rightarrow (0,1)$. These
processes was considered in Ref. \cite{two} and give, in the local optics
case, the frictional stress corresponding to the first term in (38).

In this appendix we calculate the sliding friction to linear order in the
sliding velocity when $c/\omega _p<<d<<d_W$ (where $d_W=c\hbar /k_BT$). In
this case the main contribution $\sigma $ comes from the first term in (39)
to which we must add the similar term involving $R_s$. In this integral we
replace the integration variable $q$ with $\bar p=2dq$. The integral over $%
\bar p$ is divided into two parts: the integral over $(0,\bar p_0)$ and over 
$(\bar p_0,\infty )$, where $p_0\sim dk_BT/c\hbar <<1$. In the integral over 
$(\bar p_0,\infty )$ and for $\omega >\omega _0$, where 
\[
\omega _0\sim \left( \frac c{\omega _pd}\right) ^2\frac 1\tau 
\]
we can put 
\begin{equation}
ImR_p\approx \frac 2{\bar p}\left( \frac{2\omega d}c\right) \left( \frac
\omega {2\omega _p^2\tau }\right) ^{1/2}  
\end{equation}
\begin{equation}
ImR_s\approx -\frac{2\bar pc}{\omega _pd}\left( \frac 1{\omega \tau }\right)
^{1/2}  
\end{equation}
Let us first consider the contribution $\sigma _p$ to (39) from terms
involving $R_p$. We then obtain 
\begin{eqnarray}
\sigma _p &=&\frac{\hbar v}{\pi ^2}\left( \frac 1{2d}\right) ^4\int_{\bar
p_0}^\infty d\bar p\bar p\frac d{d\bar p}\left( -\frac 1{e^{\bar
p}-1}\right) \int_0^\infty d\omega \frac \omega {2\omega _p^2\tau }\left( 
\frac{2\omega d}c\right) ^2\left( -\frac{dn}{d\omega }\right)  \nonumber \\
&\approx &\frac 3{8\pi ^2}\frac{\hbar v}{d^4}\left( 1+\int_{\bar p_0}^1d\bar
p\frac 1{e^{\bar p}-1}+\int_1^\infty d\bar pe^{-\bar p}\right) \left( \frac{%
k_BTd}{\hbar c}\right) ^2\frac{k_BT}{\hbar \omega _p}\frac 1{\omega _p\tau
}\int_0^\infty \frac{dxx^2}{e^x-1}  \nonumber \\
&\approx &\frac{3\xi }{\pi ^2}\frac{\hbar v}{d^4}\left( \frac{k_BTd}{\hbar c}%
\right) ^2\frac{k_BT}{\hbar \omega _p}\frac 1{\omega _p\tau }\left( 1+\frac
1e+\ln \frac{d_w}d\right)  
\end{eqnarray}
where $\xi =0.5986$ (see Sec. 3). The integral over $0<\bar p<\bar p_0$ can
be shown to give a negligible contribution to the linear (in the sliding
velocity) friction force. This follows from the following equation 
\[
\frac d{d\omega }\frac{\left( \mathrm{Im}R_p\right) ^2}{\mid 1-e^{-\bar
p}R_pR_p\mid ^2}\approx \frac d{d\omega }\frac{\left( \mathrm{Im}R_p\right)
^2}{\mid 1-R_pR_p\mid ^2}=\frac d{d\omega }\left( \frac{\left( \mathrm{Im}%
(s/\varepsilon )\right) ^2}{\mid s/\varepsilon \mid ^2}\right) \approx 0 
\]
where we have used that $\mathrm{Im}(s/\epsilon )/\mid s/\epsilon \mid $ is
approximately independent of frequency. Next, let us consider the
contribution $\sigma _s$ to (39) from the term involving $R_s$. We get 
\begin{eqnarray*}
\sigma _s &\approx &\frac 5{16\pi ^2}\frac{\hbar v}{d^4}\frac \hbar
{k_BT\tau }\left( \frac c{\omega _pd}\right) ^2\int_0^\infty \frac{d\bar
p\bar p^4}{e^{\bar p}-1}\int_{x_0}^\infty \frac{dx}x\frac d{d\omega }\left(
-\frac 1{e^x-1}\right) \\
&=&C\frac{\hbar v}{d^4}\left( \frac{k_BT\tau }\hbar \right) \left( \frac{%
\omega _pd}c\right) ^2
\end{eqnarray*}
where 
\[
C=\frac 5{32\pi ^2}\int_0^\infty \frac{d\bar p\bar p^4}{e^{\bar p}-1}=0.394 
\]
\[
x_0=\frac{\hbar \omega _0}{k_BT}\approx \frac \hbar {k_BT\tau }\left( \frac
c{\omega _pd}\right) ^2<<1 
\]
.

\end{document}